# In-plane polar domains enhanced energy storage

Yu Lei[1], Xiaoming Shi[2], Sihan Yan[3], Qinghua Zhang[4], Jiecheng Liu[1], Sixu Wang[5], Yu Chen[6], Jiaou Wang[6], He Qi[11], Qian Li[5], Ting Lin[7], Jingfen Li[5], Qing Zhu[8], Haoyu Wang[2], Jing Chen[1,4], Lincong Shu[9], Linkun Wang[10], Han Wu[1*], Xianran Xing[1*]

[1] *Beijing Advanced Innovation Center for Materials Genome Engineering, Institute of Solid State Chemistry, University of Science and Technology Beijing, Beijing 100083, China*

[2] *Department of Physics, University of Science and Technology Beijing, Beijing 100083, China*

[3] *College of Integrated Circuit Science and Engineering, Nanjing University of Posts and Telecommunications, Nanjing 210023, China*

[4] *Beijing National Laboratory for Condensed Matter Physics, Institute of Physics, Chinese Academy of Sciences, Beijing 100190, China*

[5] *State Key Laboratory of New Ceramics and Fine Processing, School of Materials Science and Engineering, Tsinghua University, 100084 Beijing, China*

[6] *Institute of High Energy Physics, Chinese Academy of Sciences, Beijing 100049, China*

[7] *Beijing National Center for Electron Microscopy and Laboratory of Advanced Materials, School of Materials Science and Engineering, Tsinghua University, Beijing, 100084, China*

[8] *Institute for Carbon Neutrality, University of Science and Technology Beijing, Beijing 100083, China*

[9] *College of Physics, MIIT Key Laboratory of Aerospace Information Materials and Physics, Key Laboratory for Intelligent Nano Materials and Devices, Nanjing University of Aeronautics and Astronautics, Nanjing, 211106, China*

[10] *Department of Materials Science and Engineering, Southern University of Science and Technology, Shenzhen, People's Republic of China*

[11] *Department of Physical Chemistry and Beijing Advanced Innovation Center for Materials Genome Engineering, University of Science and Technology Beijing; Beijing 100083, China*

*\* Correspondence: hanwu@ustb.edu.cn, xing@ustb.edu.cn*

*Yu Lei, Xiaoming Shi and Sihan Yan contributed equally to the manuscript.*

## Abstract

Relaxor ferroelectric thin films are recognized for their ultrahigh power density, rendering them highly promising for energy storage applications in electrical and electronic systems. However, achieving high energy storage performance with chemically homogeneous, environmentally friendly and compositionally stable materials remains challenging. In this work, we present a design of dielectrics with high energy storage performance via an in-plane polar domains incorporating polar nanoregions mechanism. Guided by phase-field simulations, we synthesized La/Si co-doping $BaTiO_3$ solid-solution thin films with high chemical homogeneity to realize high energy storage performance. Given that, we achieve a high energy density of 203.7 $J/cm^3$ and an energy efficiency of approximately 80% at an electric field of 6.15 MV/cm. This mechanism holds significant promise for the design of next-generation high-performance dielectric materials for energy storage and other advanced functional materials.

## Keywords

# Introduction

Electrostatic capacitors based on relaxor ferroelectrics (RFEs) are critical components in electronic devices and power systems due to their ultrahigh power density, rapid charge/discharge capability, high breakdown strength and reliability[1, 2, 3, 4, 5]. These advantages establish RFEs as promising candidates for energy storage applications. Significant efforts are therefore directed toward developing high-performance dielectric materials with superior energy density and efficiency to meet the miniaturization demands of advanced electronic systems[6, 7, 8]. The energy density $W_r$ of the dielectric is determined by the applied electric field $E$ and the consequent dielectric polarization $P$, $W_r = \int_{P_r}^{P_m} E\, dP$, where $P_m$ and $P_r$ are the maximum and residual polarization, respectively (Figure S1). Consequently, a combination of high $P_m$, low $P_r$, and suitable breakdown strength ($E_b$, the maximum electric field the dielectric can withstand) is imperative to attain high energy density[9].

Short-range ordered polar nanoregions enables enhanced $\Delta P$ ($=P_m-P_r$) and $E_b$, and reducd hysteresis loss[10]. Previous researchers have achieved outstanding performance. For example: (1) Yuanhua Lin et al. obtained $W_r$ = 182 J/cm$^3$ and η = 78% in Bi$_2$Ti$_2$O$_7$ thin films[11]; (2) Jingfeng Li et al. achieved $W_r$ = 202 J/cm$^3$ and η = 79% in Bi(Mg$_{0.5}$Ti$_{0.5}$)O$_3$-SrTiO$_3$ thin films[12]; (3) Weiwei Li et al. achieved $W_r$ = 215.8 J/cm$^3$ and η = 80.7% in PbZr$_{0.53}$Ti$_{0.47}$O$_3$-MgO thin films[13]. However, environmentally harmful and industrial incompatible elements limits their practical applications to a large extent[14]. Furthermore, the low-permittivity paraelectric phases, like heterogeneous layers or amorphous regions, causing generally the thin films to exhibit performance inhomogeneity and uneven morphology in nanoscale[19, 20]. This restricts operational feasibility in nanoelectronic devices.

Lead-free materials, like BaTiO$_3$ or HfO$_2$, are preferred. Sang-Hoon Bae et al. achieved $W_r$ = 191 J/cm$^3$ and η = 90% in MoS$_2$/BaTiO$_3$/MoS$_2$ heterojunctions by controlling the relaxation time using two-dimensional materials[15]. However, the complex multi-step fabrication process and the utilization of environment-sensitive two-dimensional materials pose significant challenges for scalable manufacturing and high-temperature stability, respectively[15, 16, 17].

Herein, we propose an in-plane polar domains incorporating polar nanoregions (IP-PNR) design using La/Si co-doping in $BaTiO_3$ (LSBT) thin films. Completely distinct from the mechanism of nano-sizing polar domains with heterogeneous regions to attenuate polarization, co-occupancy of La and Si with different ionic radii induces the octahedral distortion in perovskite lattice, resulting in the formation of large-sized in-plane polar domains accompanied by polar nanodomains in RFE thin films. Such a method prevents the suppression of the intrinsic polarization in lattice to a large extent, and simultaneously maintains the lower intrinsic out-of-plane polarization component. This configuration enables delayed polarization saturation while maintaining high $P_m$ at a relatively low field strength. Implementing this design in $BaTiO_3$-based thin films has yielded breakthrough performance—an ultrahigh energy density of 203.7 $J/cm^3$ with 80% efficiency at 6.15 MV/cm, setting a new benchmark for chemically homogeneous and environmentally sustainable energy storage materials.

## Results and Discussion

The IP-PNR design mechanism comprises two critical steps: (1) Controlled gradient strain engineering to disrupt long-range ferroelectric ordering, creating an out-of-plane polarization gradient that induces relaxor behavior[21], followed by (2) deliberate in-plane strain engineering via doping to reconfigure nonpolar domains surrounding PNRs into in-plane-oriented polar domains[22, 23](Figure 1a). This unique domain architecture, with polarization vectors aligned perpendicular to the applied electric field, demonstrably delays polarization saturation[24] while preserving PNRs—collectively enabling both high $P_m$ and relaxor characteristics. The combined effects of nanodomain size reduction and in-plane large-size ferroelectric domain formation synergistically enhance $\Delta P$ and $E_b$, ultimately yielding superior energy storage performance. Our materials selection focused on environmentally stable BTO as the host matrix, where precise control of growth parameters and thickness optimization enables the transformation from conventional ferroelectric to relaxor behavior. Strategic co-doping with La/Si further modulates polarization orientation within the BTO lattice.

Our study began with phase-field simulations of domain structures in designed solid solutions (Computational process see "Materials and Methods" section). The simulations revealed that undoped BTO (BTLS0) with nanoscale domains showed reduced P-E loops rectangularity and increased $\Delta P$ after strain introduction, though at the cost of decreased $P_m$. In contrast, the La/Si (LS) doped BTO system developed predominant in-plane polar T-domains along with a minor fraction of nanoscale R-domains and out-of-plane T-domains. This unique domain configuration produced P-E loops with maintained $P_r$ but significantly enhanced $P_m$. The abundance of in-plane polar domains promoted reversible domain switching[25, 26, 27], enabling electric-field-responsive polarization reorientation and delayed saturation polarization (Figure 1b and S2).

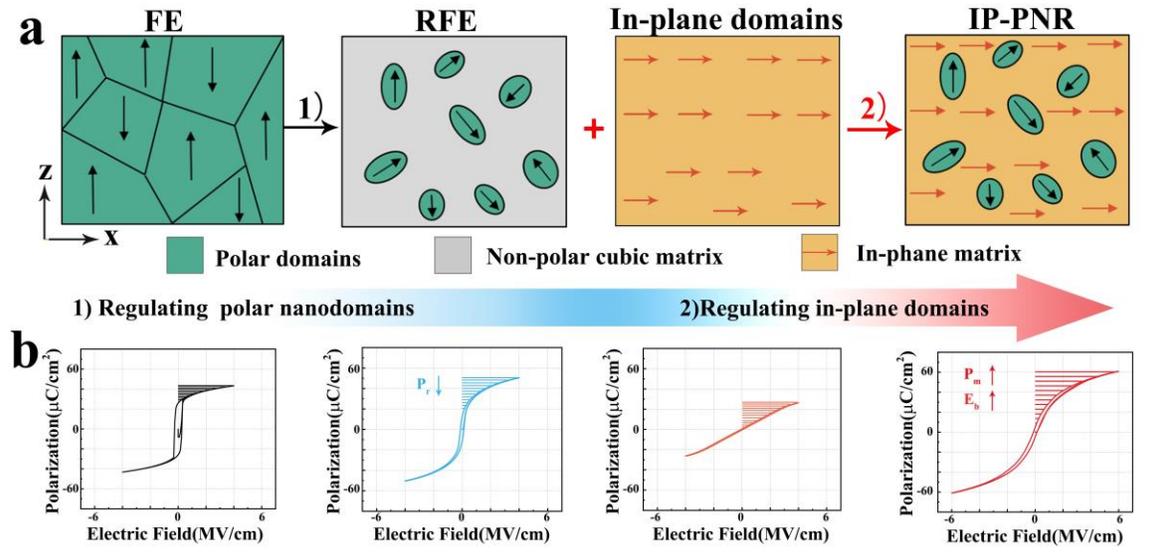

**Figure 1. Design of RFEs with enhanced energy storage performance using a ferroelectric domain surface internalization mechanism.** (**a**) Schematic diagram of domain structure (**b**) Phase field simulation of P-E loops for out-of-plane FE with micrometer domains, RFE with PNR, in-plane domain and-IP-PNR. The shaded areas in the P-E loops are indicative of the energy density.

In the present work, $BaTiO_3$-$La_2Si_2O_7$ (BTLS) thin films were grown on (100) $SrTiO_3$ substrates with $SrRuO_3$ (SRO) as conductive layers using radio frequency magnetron co-sputtering method with dual targets (Figure S5). The elemental composition of each sample was precisely quantified by Rutherford backscattering spectrometry (RBS), revealing a controlled gradient of La/Si (LS) doping concentrations (Figure S6 and Table S2). The BTLS thin films with varying doping levels were systematically labeled according to their chemical formulae: BTLS0 ($BaTiO_3$), BTLS1 ($Ba_{0.75}Ti_{0.93}La_{0.2}Si_{0.26}O_{2.88}$), BTLS2 ($Ba_{0.67}Ti_{0.88}La_{0.34}Si_{0.37}O_{2.78}$), BTLS3 ($Ba_{0.65}Ti_{0.87}La_{0.36}Si_{0.39}O_{2.73}$), BTLS4

($Ba_{0.58}Ti_{0.85}La_{0.42}Si_{0.42}O_{2.53}$). A clear correlation was observed between the sputtering power ratio and the resulting elemental composition, enabling precise control of dopant incorporation during growth (Figure S7). The thicknesses of the BTLS thin films were all around 230 nm (Figure S8). All of the thin films were verified by synchrotron radiation X-ray diffraction (SXRD) (Figure 2a and Figure S9). The ionic radius of Si is smaller than that of both Ba Ti, and the entry of Si into the lattice inevitably causes lattice contraction, leading to the peak shift towards higher angles. The occupancy of La/Si elements are determined by subsequent atomic-scale electron energy loss spectroscopy (EELS) mapping analysis. The diffraction peak intensity of thin films first increases and then decreases from BTLS1 to BTLS4 thin films, which could be ascribed to that within a certain concentration range, the increase in LS content leads to the formation of large in-plane domains, which improves the lattice ordering. Subsequently, as the LS concentration further increases, the original lattice ordering is disrupted, leading to the formation of an amorphous phase and a decrease in diffraction peak intensity[12, 28]. Furthermore, an excessive amount of LS doping would result in the emergence of impurity phases[29, 30].

The P-E loop was measured at an electric field of 4.3 MV/cm to characterize the polarization and energy storage properties (Figure 2b). The slender P-E loop exhibited classical RFE behavior, as confirmed by the broad dielectric peaks with dispersive maxima in the temperature-dependent dielectric constant measurements[4] (Figure S10). The $P_m$ value exhibited an improvement from 50 $\mu C/cm^2$ (BTLS0) to 58 $\mu C/cm^2$ (BTLS2). This enhancement is associated with the formation of large-sized in-plane polar domains, further corroborated by subsequent STEM and SHG mapping analyses. Concurrently, the $P_r$ value undergoes a progressive decline to 9 $\mu C/cm^2$. Notably, in BTLS4 thin film, which possesses a higher LS content, $P_r$ remains largely constant while $P_m$ diminishes to 47 $\mu C/cm^2$. This observation underscores the non-linear nature of the relationship between LS and the polarization characteristics of the thin films. BTLS2 thin films demonstrated superior energy storage performance at equivalent electric fields, compared with other samples.

To comprehensively assess the energy storage capacity of diverse thin films, we investigated the statistical breakdown strength $E_b$ of BTLS thin films with a Weibull distribution fitting (Figure S11a). This analysis revealed that $E_b$ attained an optimal value of 6.15 MV/cm in BTLS2 thin films, which is associated with the grain size[31]. Furthermore,

BTLS2 thin films exhibited enhanced surface flatness and grain size (Figure S12). Flat surfaces and large grain size contribute to preventing the formation of conductive channels and enhancing the breakdown field strength. Leakage current test results also corroborate the variation of breakdown field strengths (Figure S11b). LS doping widens the thin film bandgap and is second reason for the larger $E_b$ of BTLS2 thin films (Figure S13). The leakage current phenomenon is attributed to the Schottky emission mechanism, especially at high electric fields, where carrier injection is achieved by overcoming the Schottky barrier (Figure S14). It is imperative to note that the breakdown field strength exerts a substantial influence on the energy storage performance of these thin films[32]. Unipolar P-E loops analysis up to respective $E_b$ values showed BTLS2 thin film achieves exceptional $P_m$ (94 μC/cm$^2$) with minimal $P_r$, attributed to its unique IP-PNR RFE architecture (Figures S15). The simultaneous increase of $\Delta P$ and $E_b$ indicates that the maximum $W_r$ of the BTLS2 thin film is 203.7 J/cm$^3$, representing a 2-times enhancement compared to BTLS0 thin films, while maintaining a high efficiency of about 80% at an electric field of 6.15 MV/cm (Figures 2c and S16). The quality factor $W_F$ ($W_F=W_r/(100\%-\eta)$) of BTLS2 thin films reached 1018.5. Meanwhile, the maximum $W_F$ (~657) of BTLS3 thin films corresponds to a $W_r$ of 145 J/cm$^3$, which is 38% higher compared to BTLS0 thin films, and the efficiency is maintained at 79%. The energy storage performance demonstrated herein exhibits superiority over existing BaTiO$_3$-based RFE thin films. It is comparable to the best existing Pb-based and Bi-based thin films (Figure 2d, S11 a and Table S1). Furthermore, the comprehensive performance study conducted indicates that the IP-PNR RFE thin film exhibits enhanced $W_r$, even higher than that of Pb-based and Bi-based REF thin films at the same field strength, which means lower power consumption (Figure S11 b).

In the context of practical dielectric capacitor applications, the reliability and stability of energy storage performance are of paramount importance. In accelerated charge/discharge experiments with a delta electric field of 4.3 MV/cm at a frequency of 100 kHz, it was observed that the BTLS thin films all exceeded more than $1\times10^7$ cycles, whereas the BTLS2 thin films had more than $1 \times 10^9$ cycles with only a small performance degradation ($\Delta W_r$ ~ 5%, $\Delta\eta$ ~ 8%) (Figures 2e, S17 and S18). The BTLS2 thin film also exhibited good reliability at a high electric field of 5 MV/cm, sustaining more than $1\times10^8$ cycles (Figure S19). Such superior fatigue performance was comparable to these RFE thin films with high-resistivity heterogeneous content and attributed to relatively less defective thin film quality and suppressed defect migration[33, 34]. Furthermore, an investigation was conducted into their thermal stability over the temperature range of 25 to 220 °C (Figures 2f, S20 and S21). It was observed that the BTLS2 thin films exhibited superior performance in comparison to the other BTLS thin films, demonstrating only a 5% change in $W_r$ and η within the same temperature range. This enhanced cyclability and thermal stability facilitates the operation of the capacitors over extended periods and at elevated temperatures[6]. The variation of the dielectric properties of BTLS2 thin films with temperature was measured (Figure S10), and the temperature ($T_m$) of the maximum dielectric constant decreased with increasing frequency. These findings suggest that the BTLS2 thin films have enhanced relaxation properties, which improves the thermal stability of the dielectric properties. The addition of LS has also been shown to suppress the dielectric loss tangent over a wide temperature and frequency range. This behavior is associated with enhanced relaxation properties that facilitate domain switching and reduce energy dissipation[35]. Compared to the BTLS0 film, the BTLS2 film exhibits lower sensitivity of the dielectric constant to temperature variation, which can be attributed to the quasi-linear dielectric behavior induced by the in-plane polar domains[36]. The low sensitivity of the dielectric constant to temperature variation enhances the high-temperature stability of the capacitor, which is consistent with the results of the temperature stability tests. Hence, this design of IP-PNR is advantageous and effective in developing high-performance energy storage dielectrics.

Furthermore, the thickness-dependent properties of the BTLS2 thin film were evaluated (Figure S22). The results show that the optimal thickness is around 230 nm. We then measured the overdamped discharges using a 20 kΩ load resistor at 5 MV/cm (Figure S23). The BTLS2 thin film has an ultra-high discharge energy density ($W_D$) of up to 140 J/cm$^3$ and a fast discharge rate of $\tau_{0.9}$ to 2.3 μs, which shows a good charging and discharging performance, making it a candidate for pulsed power applications.

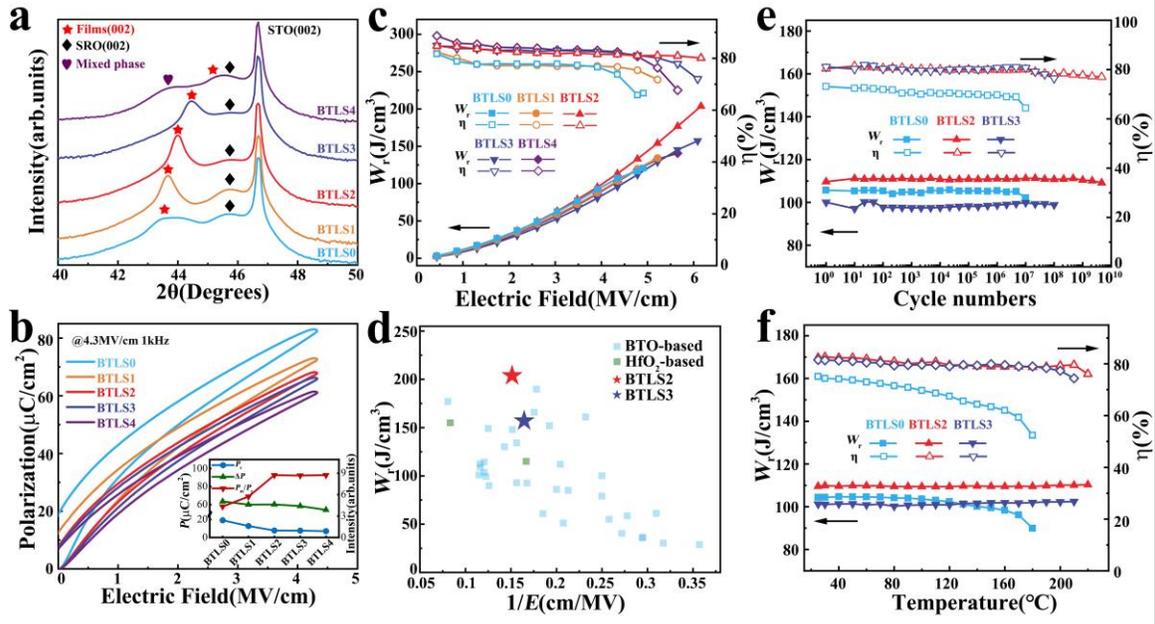

**Figure 2. Macroscopic crystal structure analysis and energy storage performance characterization of BTLS thin films.** (**a**) Synchrotron-based XRD patterns. (**b**) Unipolar P-E loops at a field strength of 4.3 MV/cm and a frequency of 1 kHz. (**c**) Energy density ($W_r$) versus efficiency ($\eta$) trajectories under increasing electric fields. (**d**) Comparison of the energy density of BTLS thin films with the BaTiO$_3$-based RFE thin films in previous studies. (**e**) Charging-discharging cycling reliability tests of BTLS0, BTLS2 and BTLS3 thin films at a field strength of 4.3 MV/cm. (**f**) Temperature stability tests of BTLS0, BTLS2 and BTLS3 thin films from 25-220 °C at a field strength of 4.3 MV/cm.

The variations in lattice ordering of perovskite-structured thin films and the effects of doping-induced oxy-octahedral distortion could be revealed by the reciprocal space mapping (RSM) analysis focused on the (103) plane. The (103) reflection of the BTLS2 thin film by azimuthally rotating the sample by 90° with respect to the surface's normal were conducted to investigate the degrees of lattice distortion of the thin films (Figure 3a). The peak position of the (013) and (0-13) reciprocal lattice reflection of the BTLS2 thin film is shifted downward (or upward) along the l-direction with respect to the (103) and (-103) reflections. This result indicates that the symmetry of the BTLS2 thin film demonstrates a symmetric monoclinically distorted lattice structure (*Pm*) induced by oxygen octahedral tilting (Table S3). Moreover, it is obvious that the (103) reciprocal lattice reflection demonstrate a dispersion degree that initially weakens and subsequently intensifies, which is consistent with the XRD results that the lattice ordering of thin films initially increase and subsequently decrease, with the variation of LS concentration (Figure S25).

The atomic-level microstructure of BTLS0 and BTLS2 thin films were characterized using a high-angle annular dark field scanning transmission electron microscopy (HAADF-STEM). The domain structure was determined based on the projected displacement of the lattice center of the B-site cation relative to its four nearest neighboring A-site cations [37,38]. STEM and geometric phase analysis (GPA) at the film-substrate interface shows that the films have good epitaxial relationships (Figure S26 and S27)[39]. Our observations indicate that the majority of the B-site cation displacement vectors of the BTLS2 (IP-PNR) thin film are directed towards the in-plane <010>, with the pointing towards the out-of-plane <001> T-distortion domain regions marked by the blue arrows and the pointing towards the diagonal <111> R-distortion domain regions marked by the green arrows (Figure 3e). Furthermore, selected-area electron diffraction (SAED) analysis also reveals the presence of monoclinic-like distortion in the thin film, which results in the polarization vectors predominantly along the in-plane orientation (Figure S28). The displacement vectors of the B-site cations pointing in in-plane direction provides the thin film with a large number of in-plane polar domains with a polarization direction reversible from in-plane to out-of-plane under applied bias. Its synergistic interaction with PNR leads to an increase in $P_m$ and a decrease in $P_r$, along with saturation polarization delay. This STEM results align closely with the phase-field simulations, indicating a high degree of consistency between the experimental setup and the theoretical models. However, the BTLS0 (PNR) thin film domain structure shows a large number of nonpolar regions in the cubic structure as well as large size polar nanoregions (Figure 3d). The polar nanoregions also contain out-of-plane <001> T-distortion domains marked by blue arrows and diagonal <111> R-distortion domains marked by yellow arrows, which are consistent with the domain structure characteristics of classical relaxor ferroelectrics. The nonpolar region could not contribute high polarization, resulting in a weaker $P_m$ for BTLS0 than for BTLS2 thin films. Different BTLS thin films require different applied electric fields to achieve the same polarization, thin films with IP-PNR RFE require larger applied electric fields due to the different flip paths of different domains(Figure S29)[40].

Extracting the B-atom displacement vectors in Figure 3e, and counting the angle and displacement magnitude shows that most of the displacement direction of B-site cations are angularly deviated from the horizontal direction by less than 15 ° in the BTLS2 thin film, and it can be assumed that the polarization of most of the polar domains pointing to in-plane direction (Figure 3h). The axial ratio mapping obtained from Figure 3e demonstrates the BTLS2 thin film possesses large-scale regions with an c/a ratio less than one, consistent with the large-size in-plane domain structures revealed in Figure 3e (Figure 3i). The axial ratio mapping is consistent with the variation of the polarization vector, demonstrating the effect of lattice strain on the polarization vector. The same domain structure can be extracted from HAADF STEM maps of different regions of the BTLS2 thin film (Figure S30). However, the BTLS0 thin film demonstrated more nonpolar domains with out-of-plane displacement directions of the B-site cations in its polar domain regions and the a/c ratio was less than one in most regions, indicating that it has fewer in-plane domain structure (Figure 3f and 3g). The angle distribution histogram extracted indicates that the in-plane domain of the BTLS2 thin film is significantly larger than that of the BTLS0 thin film (Figure S31).

We also conducted second harmonic generation (SHG) measurements with a scanning area of 10 μm ×10 μm on the BTLS0 and BTLS2 thin films (Figure 3b and 3c). The SHG intensity is proportional to the square of the in-plane polarization value (intensity∝P$_{\text{in-plane}}^2$)[41, 42]. The SHG mapping reveals an enhanced intensity distribution in the BTLS2 (IP-PNR) thin film, corresponding to the formation of in-plane strong polarization, which is consistent with the larger displacement of B-site cations and in-plane tensile strain of the BTLS2 thin film shown in Figure 3h and 3i. BTLS0 thin films, on the other hand, exhibits only locally strong polarization and a significantly lower polarization strength, indicating it has a more dispersed region of polar structural domains and also contains a high density of weakly polar regions, resulting in a relatively low $P_\text{m}$ value. Angle-resolved SHG (s-out/α=φ modes) further confirmed the changes in in-plane polarization strength as the LS content varies[43] (Figure S32).

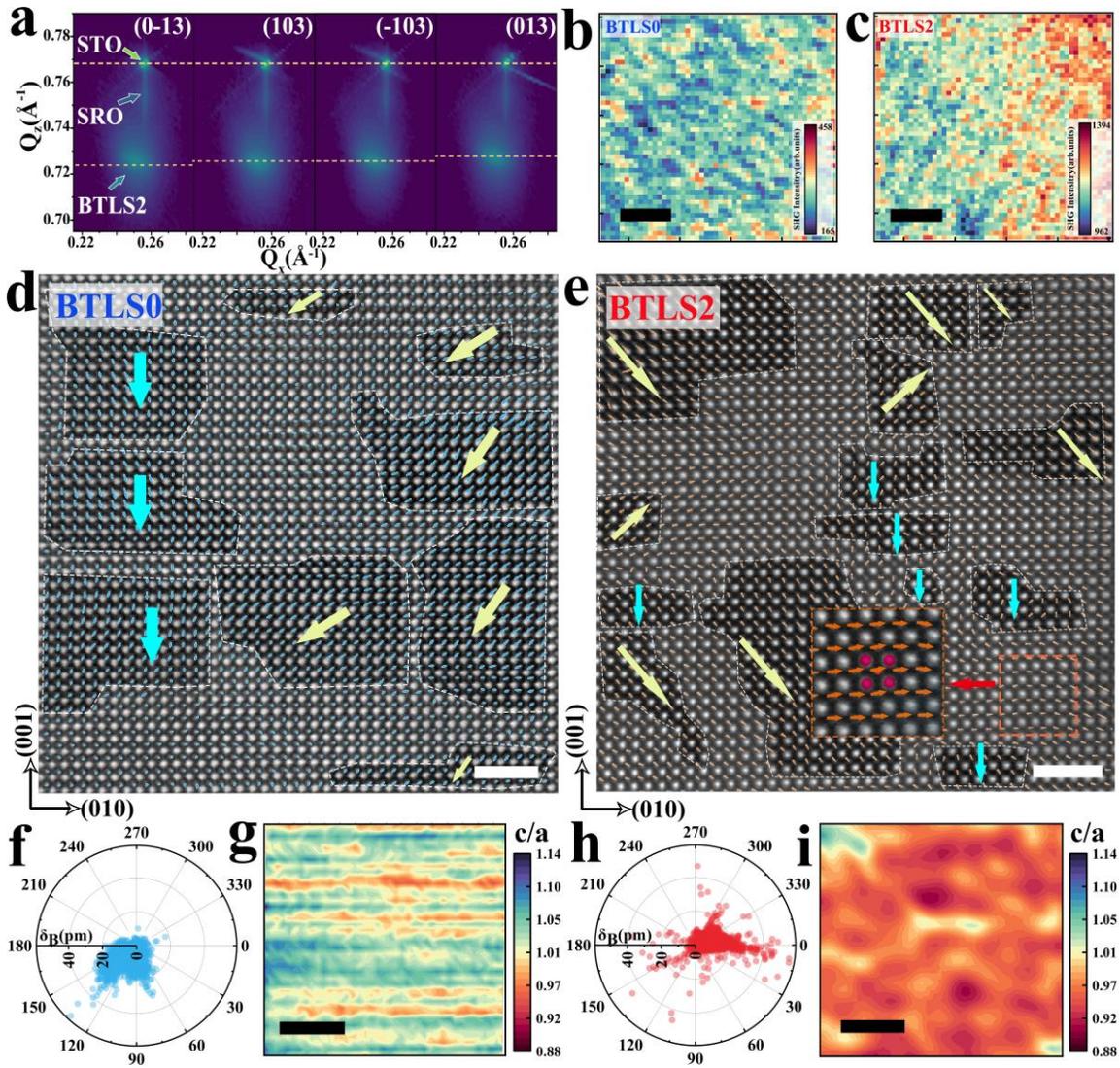

**Figure 3. Multimodal structural characterization of BTLS thin films. a**) X-ray reciprocal space mapping (RSM) images around the STO (103) reflection for the SRO/BTLS2 heterostructure. RSMs are recorded by azimuthally rotating the sample with a step size of 90° with respect to the surface's normal. (**b, c**) Second-harmonic generation (SHG) mapping of BTLS0 (PNR) and BTLS2 (IP-PNR) thin films. Scale bar, 2 μm. (**d, e**) Atomic-resolution HAADF-STEM images of BTLS0 (PNR) and BTLS2 (IP-PNR) thin films along [100] zone axe, where the overlaid polar vectors indicate the off-centering movement of B-site cations. Scale bars, 2 nm. The inset in (e) is magnified view of orange dashed-box region. (**f, g**) Quantitative analyses of polarization vector distributions and c/a-axis ratio variations for BTLS0 (PNR) thin films. Scale bars in (g), 4 nm. (**h, i**) Quantitative analyses of polarization vector distributions and a/c-axis ratio variations for BTLS2 (IP-PNR) thin films. Scale bars in (i), 4 nm.

In order to verify the homogeneity of the co-sputtered thin film growth, we conducted low-magnification STEM analysis and found no significant heterogeneous regions (Figure 4a). Minor contrast variations were due to unavoidable thickness fluctuations during FIB thinning. Furthermore, large-scale energy dispersive X-ray spectroscopy (EDXS) was utilized to show that no obvious element segregation was observed (Figure 4b). EELS mapping indicates that La occupied the A site, while Si occupied in both the A and B sites (Figure 4c). The EELS mapping also showed that all elements exhibit relatively uniform intensity, indicating that the atomic occupancy of each element corresponds to a relatively consistent atomic density. The results from the large-area EDS mapping and atomic-level element site EELS mapping consistently demonstrate that the element distribution in the BTLS2 thin film exhibits highly uniform, which is very advantageous for use in nanoscale miniaturized devices.

Moreover, the utilization of soft X-ray absorption spectroscopy (XAS) facilitates the discernment of the correlation between local symmetry breaking and polarization vector[44]. XAS of O K-edge and Ti L-edge reveals that the transition in the hybridization state of B-site Ti atoms in conjunction with the apex oxygen atoms, results in a shift in the direction of the polarization vector (Figure 4d). O K-edge XAS showed that, the A peak of BTLS2 thin films was shifted to the right and the energy was elevated, which means the Ti-O bond length becomes shorter and the deviation of Ti atoms from the center increases[45] (Figure 4e). In addition, the A` peak gradually strengthens, indicating that the degree of octahedral distortion increases[44, 46]. Besides, the B peak steadily weakened and broadened, revealing that Ti deviates from the center, the local charge distribution was chaotic, and the polarization direction is scattered. Additionally, Ti L-edge XAS showed that the splitting cleavage energy of $L_3$ increases, the stronger the octahedral field was, and the higher the cubic symmetry was, indicating that the higher the cubic structure component was in it[47] (Figure 4f). These results indicate that BTLS2 thin films exhibits greater Ti atomic displacement and more disordered polarization distribution. Although BTLS3 thin films possesses larger oxygen octahedral distortion, its higher cubic structural content suppresses polarization. Consequently, BTLS2 thin films demonstrates relatively optimal energy storage performance. Annular bright-field STEM (ABF STEM) also observed the presence of deflections in the oxygen octahedra of the BTLS2 thin film, compared to the BTLS0 thin film, corroborating the XAS results (Figure S33). Based on the crystal structure observed in STEM, the element distribution from EELS mapping, and the element ratios obtained from Rutherford backscattering, we constructed and optimized two sets of 6×6×1 supercells for DFT calculations, reasonably tracking the intrinsic physical properties of the doped structure (Figure 4g and S34). After structural optimization, the oxygen octahedra exhibited significant tilting accompanied by the lateral displacement of B-site cations, consistent with our STEM observations. This theoretically confirmed that the inconsistent ionic radii of the elements occupying the A and B sites are the direct reason of the formation of in-plane polar domains. DFT calculations also show that LS doping theoretically widens the bandgap of BTLS2 thin films, in agreement with the experimental results (Figure S35). In addition, the weak dispersion of the band structure induced by LS doping, indicates a large effective mass of charge carriers, which hinders their mobility and was also one of the reasons for the high breakdown field of the BTLS2 thin films.

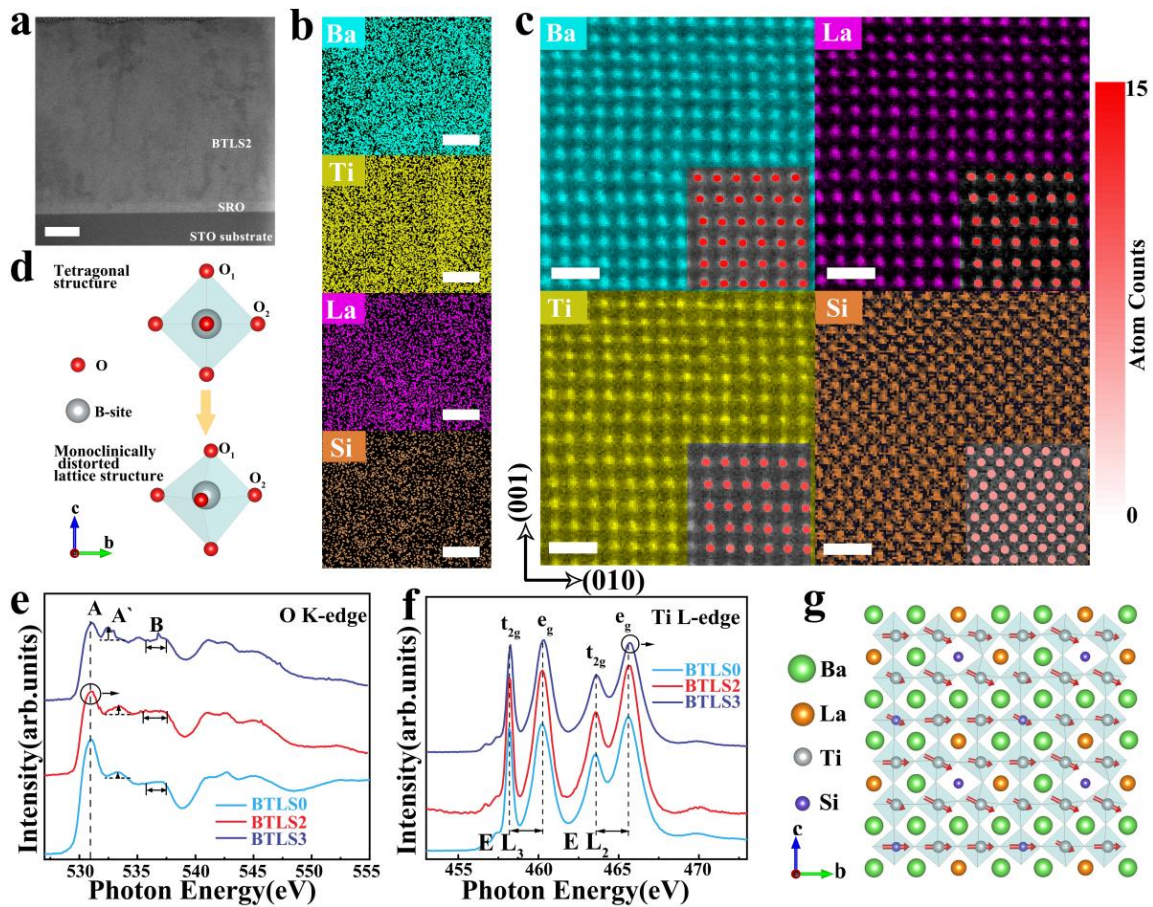

**Figure 4. Elemental distribution and site occupancy analysis, and in-plane polarization origin in thin films.** (**a**) Representative low-magnification cross-sectional STEM image of BTLS2 thin films. Scale bar, 50 nm. (**b**) EDS mapping showed the distribution of Ba, Ti, La and Si. Scale bar, 50 nm. (**c**) Atomic-resolution EELS mapping analysis of Ba, Ti, La and Si. Inset is quantified elemental column profiles. Scale bar, 1 nm. (**d**) Schematic of the deformation of the TiO$_6$ octahedron driving the rotation of the polarization direction towards the horizontal. (**e**, **f**) Experimental O K-edge and Ti L-edge spectra of BTLS0, BTLS2 and BTLS3 thin films. (**g**) DFT-optimized 6×6×1 supercell overlaid with oxygen octahedral model, wherein the shift direction of B-site (Ti/Si) cations are marked by red arrows.

## Conclusions

In summary, the IP-PNR design implemented in BaTiO$_3$-based thin films represents a breakthrough approach for modulating the intrinsic domain structure with high chemical homogeneity. Enhanced $P_m$ and reduced $P_r$, accompanied by significantly delayed polarization saturation were achieved through the complementary interaction between large in-plane polar domains and polar nanoregions. This design yields remarkable energy storage performance, reaching 203.7 J/cm$^3$ energy density with 80%, along with high stability against cycling and temperature variation. We anticipate this paradigm would inspire new research directions in developing high-performance relaxor ferroelectrics and enable novel functionalities in next-generation high-precision nanoscale energy storage device systems.

## Acknowledgments

This research was supported by National Key R&D Program of China (2020YFA0406202), National Natural Science Foundation of China. We thank the 1W1A- Diffuse X-ray Scattering Beamline of Beijing Synchrotron Radiation Facility (https://cstr.cn/31109.02.BSRF.1W1A) for providing technical support and assistance in XRD and RSM data collection. We thank the 4B9B- Photoemission Spectroscopy Beamline of Beijing Synchrotron Radiation Facility (https://cstr.cn/31109.02.BSRF.4B9B) for providing technical support and assistance in XAS data collection. We gratefully acknowledge Dr. Zhang Yi (Anhui University of Technology) for assistance with temperature-dependent dielectric measurements, and Dr. Jinglin Zhou (Beijing Synchrotron Radiation Facility, Beamline 1W1A) for synchrotron XRD characterization.